\documentclass[11pt,english,twoside]{article}

\usepackage[T1]{fontenc}
\usepackage[latin1]{inputenc}
\usepackage[english]{babel}
\usepackage{lmodern}
\usepackage{a4wide}
\usepackage{amssymb, amsmath}
\usepackage{graphicx}
\usepackage[all]{xy}
\usepackage{hyperref}
\hypersetup{linktocpage}
\usepackage{enumerate}
\usepackage{dsfont}
\usepackage{mathabx}
\usepackage{mathtools}
\usepackage{framed}

\newcommand{\beq}{\begin{equation}}
\newcommand{\eeq}{\end{equation}}
\def\bea#1\eea{\begin{align}#1\end{align}}
\newcommand{\nn}{\nonumber}
\newcommand{\w}{\wedge}

\newcommand{\ov}{\overline}
\renewcommand{\i}{\ensuremath{\textnormal{i}}}

\def\del {\partial}
\def\d {{\rm d}}
\def\mmm {\mathcal{M}}
\def\NS {N\!S}
\def\KK {K\!K}
\DeclareMathOperator{\re}{Re}
\DeclareMathOperator{\im}{Im}

\makeatletter
\@addtoreset{equation}{section}
\makeatother

\begin{document}

\begin{titlepage}

\begin{center}

\phantom{\rightline{\small AEI-2015-...}}

\vskip 3.5cm

{\fontsize{16.1}{21}\selectfont \noindent\textbf{A no-go theorem for monodromy inflation} }

\vskip 2.6cm

\textbf{David Andriot}

\vskip 0.9cm

\textit{Max-Planck-Institut f\"ur Gravitationsphysik, Albert-Einstein-Institut,\\Am M\"uhlenberg 1, 14467 Potsdam-Golm, Germany}
\vskip 0.1cm
\textit{Institut f\"ur Mathematik, Humboldt-Universit\"at zu Berlin, IRIS-Adlershof,\\Zum Gro\ss en Windkanal 6, 12489 Berlin, Germany}

\vskip 0.2cm

{\small \texttt{david.andriot@aei.mpg.de}}

\end{center}

\vskip 3.5cm

\begin{center}
{\bf Abstract}
\end{center}

\noindent We study the embedding of the monodromy inflation mechanism by E. Silverstein and A. Westphal (2008) in a concrete compactification setting. To that end, we look for an appropriate vacuum of type IIA supergravity, corresponding to the minimum of the inflaton potential. We prove a no-go theorem on the existence of such a vacuum, using ten-dimensional equations of motion. Anti-de Sitter and Minkowski vacua are ruled out; de Sitter vacua are not excluded, but have a lower bound on their cosmological constant which is too high for phenomenology.

\vfill

\end{titlepage}

\tableofcontents

\newpage

\section{Introduction}

The recent cosmological observations \cite{Ade:2013zuv, Planck:2013jfk, Ade:2013ydc, Ade:2014xna, Ade:2015tva, Ade:2015xua, Ade:2015ava, Ade:2015lrj} (see also \cite{Mortonson:2014bja, Flauger:2014qra, Adam:2014bub}) have lead an important activity on the theoretical side. The power spectrum at low $l$, the tight constraints on non-gaussianities, and the precise values or bounds on the tensor-to-scalar ratio $r$ and, for inflation, the scalar spectrum index $n_s$, are important features that a cosmological model must now respect. These experimental results have impacted the theoretical work at many levels, from general cosmological considerations to concrete models of inflation or alternatives to it, as well as embeddings of inflation in four- or ten-dimensional supergravities, or even string theory, and constraints to such models. On general aspects of cosmology, here are a few recent reviews: on the cosmological constant problem \cite{Martin:2012bt, Padilla:2015aaa}, on perturbations \cite{Chluba:2015bqa} and non-gaussianities \cite{Renaux-Petel:2015bja}. Reviews on inflation models and comparison to data can be found in \cite{Martin:2013nzq, Tsujikawa:2014rta, Clesse:2015yka, Martin:2015dha} while recent work on alternatives to it was made in \cite{Ijjas:2014nta, Ali:2014bma, Battefeld:2014uga, Ziaeepour:2014cfa, Brandenberger:2015kga, Fertig:2015dva, Quintin:2015rta}. Some classifications of inflation models were also proposed in \cite{Roest:2013fha, Wenren:2014cga}. Reviews on attempts to realise inflation in string theory can be found in \cite{Burgess:2013sla, Silverstein:2013wua, Baumann:2014nda, Chernoff:2014cba}. The tendency in supergravity or string theory is often to propose a mechanism, involving supergravity or stringy ingredients, that generates inflation in a way then tested upon the observations; in particular a single field inflaton potential is easily compared to the experimental constraints. A completion of the mechanism to a full model, with e.g. control on the various scalar fields, is however most of the time not realised; this will be the topic of this paper. Interesting examples of inflation within four-dimensional supergravities can be found in \cite{Ellis:2013nxa, Ellis:2013nka, Kodama:2015iua} (a older review is given in \cite{Yamaguchi:2011kg}); classifications of such models were proposed in \cite{Roest:2013aoa, Ceresole:2014vpa} while various constraints were pointed-out in \cite{Ferrara:2013wka, Fre:2013tya}. Embedding four-dimensional constructions in a concrete ten-dimensional compactification setting or even in string theory is not often achieved. For some constraints on doing so, see \cite{Blumenhagen:2014nba, Hebecker:2014kva, Mazumdar:2014qea, Buchmuller:2015oma, Dudas:2015lga, Das:2015uwa} as well as the recent use of the weak gravity conjecture \cite{ArkaniHamed:2006dz} and its relation to axion decay constants \cite{Banks:2003sx} (see e.g. \cite{Brown:2015iha, Junghans:2015hba, Kooner:2015rza} and references therein). Some four-dimensional examples with a possible ten-dimensional origin can be found in \cite{Blaback:2013fca, Ben-Dayan:2013fva, Czerny:2014xja, Blumenhagen:2015qda}, while some more ten-dimensional mechanisms with possible compactifications are given in \cite{Silverstein:2008sg, Marchesano:2014mla, Blumenhagen:2014gta, Hebecker:2014eua, McAllister:2014mpa, Ben-Dayan:2014lca, Abe:2014pwa, Cai:2014vua, Escobar:2015fda}. In particular, the mechanism of \cite{Kim:2004rp} (see also \cite{Kappl:2014lra}) has been embedded in several different manners in string theory \cite{Long:2014dta, Gao:2014uha, Ali:2014mra, Ruehle:2015afa, Palti:2015xra}.

In the present paper, we are interested in embedding in a concrete compactification setting the famous  monodromy inflation mechanism by E.~Silverstein and A.~Westphal \cite{Silverstein:2008sg}. This mechanism, reviewed in section \ref{sec:inflation}, considers a $D_4$-brane wrapping a circle $S^1_2$ inside a three-dimensional nilmanifold $N_3$, and moving along another circle $S^1_1$. $N_3$ is a twisted torus, i.e. a non-trivial fibration of circles. Because of this fibration, a monodromy is occurring when moving around $S^1_1$. As a consequence, the $D_4$ can travel many times around $S^1_1$ without ending-up in the same position or configuration inside $N_3$. The volume of the brane evolves monotonously with the distance travelled, so the brane moves in the direction towards the minimum of its volume. These dynamics are captured by a potential and the whole scenario can then be interpreted as an inflation mechanism: relating the distance traveled along $S^1_1$ to the inflaton $\varphi$, the potential is then in $\varphi^{\frac{2}{3}}$ for large field values and $\varphi^2$ close to the minimum. Such a potential is in good agreement with the observational constraints, making this simple mechanism very appealing. For this reason, it would be nice to embed it in a concrete compactification, where in particular global aspects are under control. This was the initial purpose of this work; however, we will rather conclude on a no-go theorem against such an embedding.

The strategy to embed the  monodromy inflation mechanism in a compactification is presented in section \ref{sec:strategy}. Because realising in one go the dynamical process looks too ambitious, the first idea is to study the static limit of the mechanism, where the brane simply stands in its vacuum configuration, or equivalently the inflaton sits at the minimum of the potential. Finding such an appropriate vacuum that contains all required elements by the mechanism is a first and necessary step in its embedding. Our study thus turns to finding such a vacuum. Among other things, it should be a vacuum of type IIA supergravity on the warp product of a four-dimensional space-time times a compact six-dimensional manifold $\mmm$. The $D_4$-brane is space-time filling, and wraps the specific internal circle inside $N_3$, the latter being contained in $\mmm$. The cancelation of the RR tadpole along the wrapped internal dimension leads us to consider a parallel orientifold $O_4$-plane. This brings further constraints because of the $O_4$ projection. Note that the backreaction of these two sources, the $D_4$ and $O_4$, is taken into account through the warp factor, even though the mechanism originally only considers the $D_4$ as a probe; this will not be in any manner constraining. The complete list of (very standard) assumptions or requirements made to find an appropriate vacuum is given in section \ref{sec:nogo}.

Before making a general search, we first focus in section \ref{sec:susy} on supersymmetric vacua with a Minkowski four-dimensional space-time. For cosmological purposes, one would rather expect the vacuum to be on de Sitter, and thus not supersymmetric. However those are notoriously difficult to find; in addition, a technique to obtain them would be to deform a supersymmetric Minkowski vacuum. So focusing on the latter in section \ref{sec:susy} is an interesting first step, and provides some intuition on the vacua to expect or shows the difficulties encountered. The tools to find supersymmetric Minkowski flux vacua on twisted tori (or solvmanifolds, see section \ref{sec:geom}) and an account on known, together with new, such vacua, were given in \cite{Andriot:2015sia}. This work is thus very useful here. We then turn in section \ref{sec:gensearch} to a more general search for an appropriate vacuum: it does not have to be supersymmetric, and the four-dimensional space-time can be anti-de Sitter, Minkowski or de Sitter. There, we essentially use the equations of motion: through various manipulations, we obtain the interesting relation \eqref{final} or \eqref{rel}. The various quantities entering it are computed making use of the assumptions, and this gives the proof of a no-go theorem in section \ref{sec:nogo}. This no-go theorem first states that anti-de Sitter and Minkowski vacua are completely excluded; secondly, a de Sitter vacuum is not excluded, but there is a lower bound on the value of its cosmological constant \eqref{finalL}, which is too high for any phenomenological purpose. This leads us to conclude negatively on any embedding of this inflation mechanism in a concrete compactification setting, at least in a phenomenologically viable manner. The reason for the absence of an appropriate vacuum was already identified in the supersymmetric case: it is discussed at the end of section \ref{sec:gensearchsusy} and in the Conclusion. Consequences for other mechanisms are also discussed there.\\

In the final stage of this project, we became aware of the interesting paper \cite{Gur-Ari:2013sba} that has some overlap with the present work. The existence of a vacuum allowing for an embedding of the inflation mechanism of \cite{Silverstein:2008sg} is studied there as well, and a no-go theorem is derived. Its conclusion is also negative for a Minkowski vacuum, or a de Sitter one with ``parametrically small'' cosmological constant: this is certainly reminiscent of our results. However, the approaches of the two papers are very different: while we only work with ten-dimensional supergravity, there the study is made in four dimensions with a scalar potential analogous to that of \cite{Silverstein:2007ac}. This necessarily induces differences, sometimes detectable (we consider localized sources, while there, they are smeared; they consider more ingredients than we do, e.g. $\NS 5$-branes and $\KK$-monopoles), sometimes less clear (is the theory considered there a consistent truncation of our ten-dimensional configurations? If yes, our result would be more general by capturing additional modes potentially allowing for other vacua). It is also unclear, although unlikely, that the ``parametrically small'' cosmological constant considered there matches the high bound obtained here. In the end, these differences amount to different assumptions, i.e. ranges of validity, of the no-go theorems; it is interesting that the conclusions remain similar.

\section{The monodromy inflation mechanism and its embedding}

\subsection{The internal geometry}\label{sec:geom}

The monodromy inflation mechanism requires a specific geometry for the internal compact manifold $\mmm$. It involves a twisted torus, also known mathematically as a solvmanifold. Such manifolds are built starting from particular Lie algebras as we now summarize. Consider a Lie algebra with vector basis $\lbrace \tilde{E}_a \rbrace$
\beq
\lbrack \tilde{E}_b , \tilde{E}_c \rbrack = \tilde{f}^a{}_{bc} \tilde{E}_a \ , \label{algebra}
\eeq
where $\tilde{f}^a{}_{bc}$ are the structure constants. One can equivalently define dual one-forms $\lbrace \tilde{e}^a \rbrace$ satisfying the relation
\beq
\d \tilde{e}^a= - \frac{1}{2} \tilde{f}^a{}_{bc} \tilde{e}^b \w \tilde{e}^c \ . \label{de}
\eeq
With the exponential map, these one-forms and relation can be promoted to the cotangent bundle of the corresponding Lie group, viewed as a manifold. A solvable algebra or group is a particular class of the above; nilpotent algebras are a further subclass included in solvable algebras. A solvmanifold $\mmm$ is made from the quotient of a solvable group $G$ by a lattice $\Gamma$: the lattice is a discrete subgroup that makes identifications (essentially of coordinates) allowing to get a compact manifold $\mmm=G/\Gamma$. The resulting manifold is also called twisted torus, since it can be viewed as fibrations of tori over tori; the fibration is encoded in the structure constants. Given a solvable algebra, the existence of a lattice is not always guaranteed, and with it the compactness; a necessary requirement for compactness is still given by the unimodularity condition $\sum_a \tilde{f}^a{}_{ab}=0$. For a review on solvmanifolds, see \cite{Bock, Andriot:2010ju}.

The one-forms $\tilde{e}^a$ of a solvmanifold are globally defined, i.e. invariant under the discrete identifications, and can be identified with the Maurer-Cartan one-forms $e^a$, although the latter are defined more generally as follows. Those are given locally in terms of the vielbein $e^a{}_m$ (with flat metric $\eta_{ab}$), as $e^a=e^a{}_m \d y^m$. The dual vector $\del_a = e^m{}_a \del_m$ matches the above $E_a$ for a solvmanifold. The quantity $f^a{}_{bc}= 2 e^a{}_m \del_{[b} e^m{}_{c]}$ matches as well the above $\tilde{f}^a{}_{bc}$ (which is constant) for a solvmanifold; with some abuse, we will call both ``structure constants''. The spin connection for the Levi-Civita connection can be expressed in terms of the $f^a{}_{bc}$, and the same holds for the Ricci tensor and scalar (see e.g. \cite{Andriot:2013xca}); one has in particular
\beq
{\cal R}=2 \eta^{ab} \del_a f^c{}_{bc} - \eta^{cd} f^a{}_{ac} f^b{}_{bd} - \frac{1}{4} \left( 2 \eta^{cd} f^a{}_{bc} f^b{}_{ad} + \eta_{ad} \eta^{be} \eta^{cg} f^a{}_{bc} f^d{}_{eg} \right)\ . \label{Ricciflat}
\eeq
This expression simplifies for a solvmanifold, using the unimodularity condition.

The monodromy inflation mechanism \cite{Silverstein:2008sg} requires a manifold $\mmm$ that contains a specific twisted torus: it is the three-dimensional nilmanifold $N_3$, sometimes called the Heisenberg manifold since it is built from the Heisenberg algebra. This translates into the following relations
\beq
\d \tilde{e}^1=0 \ ,\ \d \tilde{e}^2=0\ ,\ \d \tilde{e}^3 = -\tilde{f}^3{}_{12} \tilde{e}^1\w \tilde{e}^2 \ .\label{e123}
\eeq
For unit radii, one takes $\tilde{f}^3{}_{12}=M \in \mathbb{Z}^*$ and a local expression of the one-forms is given by
\beq
\tilde{e}^1=\d y^1\ , \ \tilde{e}^2=\d y^2\ , \ \tilde{e}^3=\d y^3 - M y^1 \d y^2 \ . \label{Heisea}
\eeq
Maintaining these one-forms invariant, one can read the discrete identifications to be made on the coordinates
\beq
(y^1,\ y^2,\ y^3) \sim (y^1,\ y^2,\ y^3 + 2\pi) \sim (y^1,\ y^2 + 2\pi,\ y^3) \sim (y^1 + 2\pi,\ y^2,\ y^3 + 2\pi M y^2) \ ,
\eeq
We see that $y^1$ and $y^2$ are well-defined coordinates for standard circles $S^1_1$ and $S^1_2$, but $y^3$ is not and the corresponding direction is twisted, i.e. non-trivially fibered over the other two directions. Indeed, consider an object located at $(y^1, y^2, y^3)$ (given the discrete identification, we take $y^2 \in \lbrack 0, 2\pi \lbrack$), and let it move once around the circle $S^1_1$. Its locus in $(y^1, y^2)$ is unchanged, but is a priori changed along the third direction. The third coordinate is shifted by $2\pi M y^2$, which can only be identified with the initial position if $M y^2 = p\in \mathbb{Z}$, i.e. for $y^2=\frac{p}{M} <2\pi$. There are therefore Integer$[2\pi M]$ positions $y^2$ such that this occurs, i.e. a finite set of initial positions, otherwise the object ends-up at a different $y^3$. Thus, most likely, the object is not located at the same point in the twisted torus after one round along $S^1_1$, so there is a monodromy in the third direction. This monodromy will be the source of the inflation mechanism now to be described.

\subsection{The monodromy inflation mechanism}\label{sec:inflation}

The monodromy inflation mechanism of \cite{Silverstein:2008sg} considers a standard split of the ten-dimensional space-time into a four-dimensional one, times a compact internal six-dimensional manifold $\mmm$. The latter should contain the three dimensional nilmanifold just described, in order to benefit from the monodromy. The mechanism considers a probe $D_4$-brane, that is space-time filling, and wrapping the circle along direction $\tilde{e}^2$ of the above twisted torus. The brane moves along the circle $S^1_1$; as we will describe, the related open string modulus, namely the distance traveled by the brane along this direction $\tilde{e}^1$, is related to the inflaton.

We just explained that an object located at a point in directions $2$ and $3$ does not come back to the same point along $3$ after going around the circle $S^1_1$, except for special loci in $2$. Since the brane is an extended object wrapping the whole $S^1_2$, it never comes back to itself, even after several rounds along $S^1_1$. Thanks to this monodromy, it can move for a ``long time'' inside the internal space, by going around $S^1_1$. Since the inflaton is related to the distance traveled along that direction, this can generate a ``large field inflation''. This is achieved without any constraint on the actual size of the internal space, on the contrary to other mechanisms considering a brane moving inside a throat.

The reason for such dynamics has not yet been spelled-out. It can be better viewed from the DBI action of this brane, that gives eventually the inflaton potential. Considering no world-volume gauge field, the action is essentially given by the determinant of the pulled-back metric. Taking for simplicity unit radii, the three-dimensional metric is locally written
\beq
\d s_3^2 = (\tilde{e}^1)^2 + (\tilde{e}^2)^2 + (\tilde{e}^3)^2 = (\d y^1)^2 + \left(1 + (M y^1)^2 \right) (\d y^2)^2 + (\d y^3)^2 -2 M y^1 \d y^2 \d y^3 \ . \label{metric3}
\eeq
The only motion considered is along $S^1_1$, so the determinant of the pulled-back metric is written \cite{Silverstein:2008sg} as
\beq
\mbox{det}P[g]= \mbox{det}\left(g_{mn} \frac{\del X^m}{\del \xi^i}\frac{\del X^n}{\del \xi^j}\right) = \mbox{det}(g_4) (1 - g_{11}(\del_t X^1)^2) g_{22} \ , \label{det}
\eeq
where $g_4$ denotes here the external four-dimensional metric, $g_{11}$ and $g_{22}$ refer to the twisted torus metric \eqref{metric3} and the square includes the absolute value of the metric time component inverted. The DBI action however involves the induced metric seen by the brane, meaning one should take the metric on the covering space and use the brane coordinates or open string moduli $X^m$ instead of $y^m$ \cite{Silverstein:2008sg}. The open string modulus $X^1$ along $S^1_1$ is related to the inflaton $\varphi$; the DBI Lagrangian, given essentially by the square root of \eqref{det}, eventually leads to the kinetic term and the potential of the inflaton. The non-trivial contribution to the potential is through $g_{22}(X^1)=1 + (M X^1)^2$. The potential is then growing when moving away from $X^1=0$. Another perspective is to consider the internal volume of the brane: it depends similarly on $g_{22}(X^1)$, and thus grows when going away from $X^1=0$. This volume being related to the energy, the brane therefore tends to relax by minimizing its volume and thus moves along $S^1_1$ to reach $X^1=0$; equivalently, the inflaton rolls down the potential to reach the minimum in $X^1=0$. This creates the dynamics.

We refer to the original work \cite{Silverstein:2008sg} for more details, but let us recall that the generated inflaton potential is in $\varphi^{\frac{2}{3}}$ for large field values, and $\varphi^2$ close to the minimum. As already mentioned, this is in good agreement with the latest observational constraints, so this mechanism is appealing. The next step is to embed it in a concrete compactification setting: we now turn to this question.

\subsection{Strategy for the embedding: the vacuum}\label{sec:strategy}

We would like to embed the monodromy inflation mechanism just described in a concrete compactification setting. Starting with the complete dynamical process might be difficult; rather a reasonable first case to consider is the static limit, where the inflaton is simply at the minimum of the potential, namely the vacuum, or equivalently the $D_4$-brane is static at $X_1=0$. This limiting case should in any case be covered by a more general embedding. So the strategy consists in finding an appropriate vacuum with the necessary features for the mechanism, and the standard properties of a compactification. We now give some details on these characteristics, while the complete list of assumptions on this vacuum is summarized in section \ref{sec:nogo}.

We will look for a vacuum of ten-dimensional type IIA supergravity with standard NSNS and RR fluxes, and sources for the latter ($D_p$-branes and orientifolds $O_p$-planes). We do not consider any $\NS5$-brane neither $\KK$-monopoles, despite the latter is advocated in \cite{Silverstein:2007ac, Silverstein:2008sg} for a de Sitter vacuum: finding an explicit vacuum with those together with the $D_4$-brane looks too difficult; see also footnote \ref{foot:KK} on including these ingredients. The ten-dimensional space-time is split as a warped product of a maximally symmetric four-dimensional space-time, along directions $\d x^{\mu}$, times a six-dimensional compact manifold $\mmm$, along directions $\d y^m$. The metric is written accordingly
\beq
\d s^2= e^{2A(y)} \tilde{g}_{\mu\nu} (x) \d x^\mu \d x^\nu + g_{mn} (y) \d y^m \d y^n \ ,\label{10dmetric}
\eeq
with warp factor $e^A$, and $\tilde{g}_{\mu\nu}$ the metric of de Sitter, Minkowski or anti-de Sitter space-time. As a general convention, the tilde metric would always denote the one without warp factor, that can be viewed as the smeared metric. The sources $D_p$ and $O_p$ will be space-time filling, and wrap part of the internal manifold. The warp factor captures their backreaction. Note that for the $D_4$-brane, this is more refined than required by the inflation mechanism, where it is only treated as a probe; this refinement will however not be constraining for finding the vacuum. Finally, the dilaton will only depend on internal coordinates, and the fluxes will be purely internal.

From a cosmological perspective, the four-dimensional space-time in the vacuum should be de Sitter; we will nevertheless make a more general search including Minkowski or anti-de Sitter. Obtaining de Sitter vacua is notoriously difficult, in particular purely perturbative and metastable ones. One technique is to start with a Minkowski or anti-de Sitter one and deform it to de Sitter by corrections, such as non-perturbative contributions or turning on fluxes, etc. Allowing as well for these two space-times in our search will then inform us on the possibilities of such uplifts to de Sitter.

The inflation mechanism imposes constraints on the vacuum. To start with, the manifold $\mmm$ should contain to some extent the above nilmanifold $N_3$. We will then consider $\mmm$ to be a solvmanifold, only deformed by warp factors, and discuss below how to include $N_3$. Another point is having the $D_4$-brane, wrapping the direction along $\tilde{e}^2$. These two points have implications that we now present.
\begin{itemize}
\item In the vacuum, the Bianchi identities of the RR fluxes should be satisfied, or equivalently the tadpole should be canceled. In presence of the $D_4$-brane, a Minkowski or de Sitter vacuum requires to have negative contributions to the tadpole \cite{Maldacena:2000mw, Giddings:2001yu}. In our context, these will be provided by orientifold planes. More precisely, $O_4$-planes should wrap as well the $\tilde{e}^2$ direction. For anti-de Sitter vacua, these orientifolds are not strictly necessary, but we consider them nevertheless in view of an uplift.

\item The nilmanifold $N_3$ cannot be contained randomly in $\mmm$. The guideline is to preserve the determinant \eqref{det} that leads to the potential of interest for the inflaton. To do so, the brane world-volume should be trivially embedded into $\mmm$ as used to obtain \eqref{det}, so we need $\tilde{e}^2=\d y^2$, i.e. the direction $2$ (wrapped) is not fibered but simply given by a coordinate. We also need $\tilde{e}^1=\d y^1$, i.e. not fibered, to define properly the coordinate along $1$, eventually appearing in $g_{22}$ and related to the inflaton. Finally, we need $\tilde{f}^3{}_{12} \neq 0$ for the mechanism, while other structure constants are in principle possible. One could however put further restrictions on the local expressions of the one-forms: additional structure constants could lead undesired contributions to the metric and the determinant \eqref{det}; but we will not need to do so. The presence of $O_4$-planes along $\tilde{e}^2$ also imposes further restrictions. The orientifold projection has to be compatible with the manifold, which translates to conditions on the algebra: preserving it under the involution $\sigma$ imposes the only non-zero structure constants to be $\tilde{f}^a{}_{b2}\neq 0$. All this eventually restricts the one-forms of the underlying solvmanifold to satisfy
\bea
& \d \tilde{e}^1=\d \tilde{e}^2=0\ , \ \d \tilde{e}^3= - \tilde{f}^3{}_{12}\, \tilde{e}^1\w \tilde{e}^2 -\sum_{b\neq1,2} \tilde{f}^3{}_{b2}\, \tilde{e}^b \w \tilde{e}^2\ , \label{manifold}\\
& \d \tilde{e}^{a=4,5,6}= -\sum_{b\neq2} \tilde{f}^a{}_{b2}\, \tilde{e}^b \w \tilde{e}^2 \ ,\qquad \tilde{f}^3{}_{12} \neq 0 \ .\nn
\eea

\end{itemize}
Let us fix further the notations. As introduced already, the one-forms $\tilde{e}^a$ denote those of the underlying solvmanifold, while $e^a=e^a{}_m \d y^m$ are those of the internal manifold $\mmm$. As for the metric with or without tilde, the difference between these one-forms is only a warp factor. We will eventually (see section \ref{sec:nogo}) restrict ourselves to the case with only $D_4$ and $O_4$ along $\tilde{e}^2$, and transverse to the other one-forms denoted accordingly $\tilde{e}_{\bot}^a$; the dependence on the warp factor is then obvious and the internal metric is block diagonal in this basis
\beq
\d s^2_6= g_{mn} \d y^m \d y^n = \eta_{ab} e^a e^b = e^{2A} (\tilde{e}^2)^2 + e^{-2A} \eta_{ab} \tilde{e}_{\bot}^a \tilde{e}_{\bot}^b \ . \label{internalmetric}
\eeq
In these notations, the $\tilde{e}^a$ may still contain some radii so the determinant of the corresponding vielbeins does not have to be $1$.

We now turn to the search of an appropriate vacuum. We first look for supersymmetric Minkowski vacua: these are simpler to find, and will provide some intuition. We will then make a more general study.

\section{Warm-up: looking for an appropriate supersymmetric Minkowski vacuum}\label{sec:susy}

\subsection{Conditions for supersymmetric Minkowski vacua}\label{sec:condsusy}

The conditions to solve, in order to get an ${\cal N}=1$ supersymmetric Minkowski vacuum (of the general form described above), have been reviewed in section 2.1 of \cite{Andriot:2015sia}, and we only present here the needed material. We use the convenient language of SU(3)$\times$SU(3) structures on $\mmm$. An SU(3)$\times$SU(3) structure of the generalized tangent bundle of generalized complex geometry is described for us by a pair of polyforms (sums of forms of different degrees) $\Phi_{\pm}$. Such polyforms verify some structure compatibility conditions. The explicit expressions of $\Phi_{\pm}$ depends on the structure group of the standard tangent bundle of $\mmm$, for which one should distinguish various cases. For constant ones, there are three possibilities: an SU(3), an orthogonal (or static) SU(2), or an intermediate SU(2) structure. Non-constant ones, such as dynamical SU(2) structure are very unlikely to be found (none is known on compact manifolds), so we do not consider them here.

Most conditions can be phrased in terms of $\Phi_{\pm}$. The Killing spinor equations can be reformulated as the following differential conditions \cite{Grana:2005sn, Grana:2006kf}
\bea
& (\d -H\w)(e^{2A-\phi}\Phi_{\pm})=0 \ , \label{SUSY}\\
& (\d -H\w)(e^{A-\phi}\re(\Phi_{\mp}))=0 \ , \nn\\
& (\d -H\w)(e^{3A-\phi}\im(\Phi_{\mp}))=\pm\frac{e^{4A}}{8}  *\lambda(F) \ , \nn
\eea
where $\phi$ is the dilaton, the upper sign is for IIA and lower for IIB, and $*$ is the internal Hodge star for which we use the standard convention as in \cite{Andriot:2015sia}.\footnote{The norm of internal spinors, or equivalently of the polyforms, has been related to the warp factor, as required in the presence of an orientifold.} $H=\d b$ is the NSNS three-form flux and the RR fluxes are captured by the $p$-forms $F_p$ gathered as
\bea
\textrm{IIA}&:\ F=F_0+F_2+F_4+F_6\ ,\ \lambda(F)=F_0-F_2+F_4-F_6 \ , \label{RRIIA} \\
\textrm{IIB}&:\ F=F_1+F_3+F_5\ ,\ \lambda(F)=F_1-F_3+F_5 \ . \label{RRIIB}
\eea
$H$ and $F_p$ are purely internal forms.\footnote{\label{foot:demo}The RR fluxes appearing in these conditions were initially defined using the democratic formalism: they are the internal forms $F^{6}_p$ defined as follows from their ten-dimensional counterparts
\beq
F^{10}= F^{6} \pm {\rm vol}_4 \w \lambda (* F^{6})
\eeq
where the polyform $F^{6}$ is the $F$ in \eqref{RRIIA} or \eqref{RRIIB}, and $F^{10}$ is a polyform defined analogously for the ten-dimensional self-dual RR fluxes of the democratic formalism; ${\rm vol}_4 $ is the warped four-dimensional volume form. It is clear that $F^{10}_0= F^{6}_0 \ , \ F^{10}_2= F^{6}_2$ while $F^{10}_4$ could get a further, external, contribution. However the supersymmetry conditions will impose $F_6^6=0$; for the more general solutions looked for, we will consider as well that $F^{10}_4$ has only internal components. So we get as well $F^{10}_4= F^{6}_4$, and we drop the upper numbers.} The properties of the sources are also encoded in terms of the polyforms. If the sources are compatible with the bulk supersymmetry, their internal world-volume form ${\rm vol}_{||}$ is given by the pullback of $\im(\Phi_{\mp})$
\beq
P[ \im(\Phi_{\mp}) ] = \frac{e^A}{8} {\rm vol}_{||} \ .\label{calib}
\eeq
In addition, if the last equation of \eqref{SUSY} is satisfied (a particular case being that the (bulk) background is supersymmetric), then the energy of these sources is minimized, and they are calibrated \cite{Koerber:2005qi, Martucci:2005ht, Koerber:2006hh, Koerber:2007hd, Witt, Martucci:2011dn}. Finally, the orientifold projection conditions \cite{Grana:2006kf} can be encoded as follows for an $O_4$ or $O_8$
\beq
\sigma(\Phi_+) = - \lambda(\Phi_+) \ ,\ \sigma(\Phi_-) = - \lambda(\ov{\Phi}_-) \ .\label{O4projPhi}
\eeq
The Bianchi identities (BI) of the fluxes are further constraints to be satisfied. This is usually difficult, because of the appearance of undesired terms, that could correspond for instance to sources along directions that are not allowed. These terms are often due to the non-trivial fibration of the underlying manifold, or to multiple warp factors for intersecting sources; we refer to \cite{Andriot:2015sia} for a discussion on those. We will not need to study in more details these BI here. Let us nevertheless recall that if the supersymmetry conditions and the BI are satisfied, the equations of motion are automatically implied \cite{Lust:2004ig, Gauntlett:2005ww, Grana:2006kf, Koerber:2007hd}.

Supersymmetric vacua with $O_4$ are actually very restricted: the projection \eqref{O4projPhi} does not allow for SU(3) or intermediate SU(2) structures \cite{Grana:2006kf, Koerber:2007hd}, so we are left with an orthogonal SU(2) structure solution. The polyforms are then given by
\beq
{\mbox SU(2)}_{\bot}: \ \Phi_+ = -\frac{e^A}{8} \i e^{\i \theta_+} \omega\w e^{\frac{1}{2} z\w \ov{z}} \ ,\ \Phi_- = -\frac{e^A}{8} e^{\i \theta_-} z\w e^{-\i j} \ ,
\eeq
where $\theta_{\pm}$ are constant phases, and $j$ is a (1,1)-form, $z$, $\omega$, are (1,0)-, (2,0)-forms, with respect to an almost complex structure $I$. The structure compatibility conditions to be verified then become
\bea
& j^2=\frac{1}{2} \omega\w \ov{\omega} \neq 0 \ ,\ j\w \omega=0\ ,\ \omega\w \omega=0 \label{compatjo}\\
& z\vee \omega= 0 \ ,\ z\vee j=0 \label{compatz} \ .
\eea
Following appendix A of \cite{Andriot:2015sia}, the supersymmetry conditions \eqref{SUSY} are now written
\bea
& \d (e^{2A} \omega)=0 \label{susy1}\\
& \omega\w \d (\frac{1}{2} z\w \ov{z})= H\w \omega \label{susy2}\\
& \d (e^{A} \re(z))=0 \label{susy3}\\
& \d (e^{A} \im(z)\w j)= e^{A} H\w \re(z) \label{susy4}\\
& \re(z)\w \d (\frac{1}{2} j^2)= H\w \im(z)\w j \label{susy5}\\
& F_6=0 \label{susy6}\\
& \d (e^{3A} \im(z))= g_s e^{4A} * F_4 \label{susy7}\\
& \re(z)\w \d (e^{2A} j)= e^{2A} H\w \im(z) - g_s e^{3A} * F_2 \label{susy8}\\
& \im(z)\w j\w \d (\frac{e^{2A}}{2} j)= -\frac{1}{2} e^{2A} H\w \re(z) \w j + g_s e^{3A} * F_0 \ , \label{susy9}
\eea
where we fixed $\theta_-=\pi$, and took for an $O_4$ $e^{\phi}=g_s e^A$ with a constant $g_s$. The supersymmetry condition for the source \eqref{calib} becomes
\beq
P[ \im (z) ] = {\rm vol}_{||} \ ,\label{calibO4}
\eeq
while the orientifold projection \eqref{O4projPhi} translates to
\beq
\sigma(\re(z))= - \re(z)\ ,\quad \sigma(\im(z))= \im(z)\ ,\quad \sigma(\omega)= \omega\ , \quad \sigma(j)=j \label{O4proj}\ .
\eeq
Finding a vacuum requires to solve the above conditions, in particular for us on a manifold of the form \eqref{manifold}.

Using the notations of \eqref{internalmetric}, we can already find the most general forms compatible with the projection \eqref{O4proj} \cite{Grana:2006kf}
\beq
\omega= z^1 \w z^2 = (e^{-A} \tau^1_a \tilde{e}^a_{\bot} ) \w (e^{-A} \tau^2_a \tilde{e}^a_{\bot} ) \ ,\ z= ( e^{-A} \tau^3_a \tilde{e}^a_{\bot} + \i e^{A} \tilde{e}^2 ) \ , \ j \approx e^{-2A} \tilde{e}^a_{\bot} \w \tilde{e}^b_{\bot} \ , \label{structforms}
\eeq
where the expression of $j$ does not need to be specified further, and all coefficients are constant. $\tau^{1,2}_a$ are complex and $|\tau^{1,2}_a|= 0 \ {\rm or}\ 1$, $\tau^3_a$ is real and equals $0 \ {\rm or}\ 1$. This normalisation should allow to reproduce the metric \eqref{internalmetric} from the structure forms and the almost complex structure, following the procedure detailed in \cite{Andriot:2015sia}. We are now going to study whether these structure forms can satisfy the supersymmetry conditions.

\subsection{A vacuum and a first no-go on $N_3 \times T^3$}\label{sec:N3T3}

We consider here a particular case of the manifolds allowed in \eqref{manifold}, which is the simple product of the nilmanifold $N_3$ times the three-torus
\beq
\d \tilde{e}^1=\d \tilde{e}^2=0\ , \ \d \tilde{e}^3= - \tilde{f}^3{}_{12}\, \tilde{e}^1\w \tilde{e}^2\ ,\ \d \tilde{e}^{4}= \d \tilde{e}^5=\d \tilde{e}^6= 0 \ ,\qquad \tilde{f}^3{}_{12} \neq 0 \ . \label{niltorus}
\eeq
We are going to prove that there exists no vacuum of the form discussed above on this manifold, the crucial point being that the sources are along $\tilde{e}^2$, giving the forms \eqref{structforms}. Indeed, before we turn to this no-go, let us first recall a vacuum that exists on this manifold with $O_4$ and $D_4$ along $\tilde{e}^3$: this will give an illustration of the expected vacua.

\subsubsection*{A vacuum with sources along $\tilde{e}^3$}

This vacuum was first obtained from one a torus with $H$-flux, by applying T-duality \cite{Kachru:2002sk}.\footnote{A further T-duality was shown to bring it to a non-geometric background, a point studied in more details in \cite{Marchesano:2007vw, Andriot:2011uh}; this vacuum thus played an important role in the non-geometry literature, and is sometimes referred to as the toroidal example. A partial quantization of the closed string was performed on this vacuum in \cite{Andriot:2012vb}.} We presented it in details in section 3.1 of \cite{Andriot:2015sia}, and only recall here the main features.

The solution has $O_4$ and $D_4$ along $\tilde{e}^3$: the solvmanifold \eqref{niltorus} is compatible with this projection. With respect to the above material, the only difference is the orthogonal SU(2) structure forms given by
\beq
\omega=z^1\w z^2=e^{-2A} (\tilde{e}^1 + \i \tilde{e}^2)\w (\tilde{e}^4 + \i \tilde{e}^5)\ ,\ z= e^{-A} e^6 + \i e^A e^3 \ ,\ j=\frac{\i}{2} (z^1\w \ov{z^1} + z^2\w \ov{z^2})
\eeq
and the corresponding metric, given by \eqref{internalmetric} with $\tilde{e}^2$ and $\tilde{e}^3$ exchanged. All above conditions are satisfied. The only non-zero flux is
\bea
F_4 =-g_s^{-1} & \left( e^{4A} *\left( \d(e^{-4A})\w \tilde{e}^3\right) + \tilde{f}^3{}_{12} \tilde{e}^3\w \tilde{e}^4\w \tilde{e}^5\w \tilde{e}^6 \right) \\
= - g_s^{-1} &\Bigg( - \eta^{11} \del_{\tilde{1}} (e^{-4A}) \tilde{e}^2\w \tilde{e}^4\w \tilde{e}^5\w \tilde{e}^6  + \eta^{22} \del_{\tilde{2}} (e^{-4A}) \tilde{e}^1\w \tilde{e}^4\w \tilde{e}^5\w \tilde{e}^6  \nn\\
&\ - \eta^{44} \del_{\tilde{4}} (e^{-4A}) \tilde{e}^1\w \tilde{e}^2\w \tilde{e}^5\w \tilde{e}^6 + \eta^{55} \del_{\tilde{5}} (e^{-4A}) \tilde{e}^1\w \tilde{e}^2\w \tilde{e}^4\w \tilde{e}^6 \nn\\
&\ - \eta^{66} \del_{\tilde{6}} (e^{-4A}) \tilde{e}^1\w \tilde{e}^2\w \tilde{e}^4\w \tilde{e}^5 + \tilde{f}^3{}_{12} \tilde{e}^3\w \tilde{e}^4\w \tilde{e}^5\w \tilde{e}^6 \Bigg) \label{F4standard}
\eea
where $A(y^1, y^2, y^4, y^5, y^6)$, and the derivatives are the $\del_{\tilde{a}}= \tilde{e}^m{}_{a} \del_m$. We then obtain
\beq
\d F_4= - g_s^{-1}  \Big( \tilde{\Delta} (e^{-4A}) + \left(\tilde{f}^3{}_{12} \right)^2 \Big) \widetilde{{\rm vol}}_{\bot} \ , \label{BIe3}
\eeq
in terms of the smeared transverse volume to the source $\widetilde{{\rm vol}}_{\bot}= -  \tilde{e}^1\w \tilde{e}^2\w \tilde{e}^4\w \tilde{e}^5\w \tilde{e}^6 $. The expression \eqref{BIe3} gives the expected BI when equating the source contributions. As discussed in \cite{Andriot:2015sia}, the constant term in $\left(\tilde{f}^3{}_{12} \right)^2$ is a standard one, contributing to the tadpole cancelation, and shifting the source contributions (more $O_4$ or less $D_4$).

\subsubsection*{No vacuum with sources along $\tilde{e}^2$}

We now prove that there exists no appropriate vacuum on \eqref{niltorus} given the forms \eqref{structforms}. To do so, we first focus on the conditions \eqref{susy1} and \eqref{susy3}
\bea
& \d(e^A \re z)=0 \ , \label{susyz}\\
& \d(e^{2A} \omega)=0 \ . \label{susyo}
\eea
We first consider \eqref{susyz} and obtain
\beq
\tau^3_3  \tilde{f}^3{}_{12}\, \tilde{e}^1\w \tilde{e}^2 =0 \Rightarrow \tau^3_3 = 0 \ . \label{result1}
\eeq
The general form of $z$ is thus modified, as well as $z \w \overline{z}$
\beq
z= e^{-A} \sum_{a=1,4,5,6} \tau^3_a \tilde{e}^a  + \i e^{A} \tilde{e}^2  \ ,\quad z \w \overline{z} = 2 \i \sum_{a=1,4,5,6} \tau^3_a  \tilde{e}^2 \w \tilde{e}^a \ . \label{zovz}
\eeq
We now consider \eqref{susyo} and get similarly
\bea
& \sum_{a,b= \bot} \tau^1_{[a} \tau^2_{b]} \sum_{c,d} \tilde{f}^a{}_{cd}\, \tilde{e}^b \w \tilde{e}^c \w \tilde{e}^d = 0 \ ,\\
\Leftrightarrow & \sum_{b= 4,5,6} \tau^1_{[3} \tau^2_{b]} \tilde{f}^3{}_{12}\, \tilde{e}^b \w \tilde{e}^1 \w \tilde{e}^2 = 0 \ ,\\
\Leftrightarrow & \forall b= 4,5,6,\ \tau^1_{[3} \tau^2_{b]} =0 \ .\label{result2}
\eea
We deduce the following modifications
\bea
\omega= e^{-2A} \Bigg(& 2 \tau^1_{[1} \tau^2_{3]} \tilde{e}^1\w \tilde{e}^3 + \sum_{a,b=1,4,5,6} \tau^1_{[a} \tau^2_{b]} \tilde{e}^a\w \tilde{e}^b \Bigg) \ , \label{ome}\\
\omega \w \overline{\omega} = e^{-4A} \Bigg(& 4 \sum_{a,b=4,5,6} \re\left(\tau^1_{[1} \tau^2_{3]}\, \overline{\tau^1_{[a}} \overline{\tau^2_{b]}}\right) \tilde{e}^1\w \tilde{e}^3 \w \tilde{e}^a\w \tilde{e}^b \label{oovo}\\
& + \sum_{a,b,c,d=1,4,5,6} \tau^1_{[a} \tau^2_{b]}\, \overline{\tau^1_{[c}} \overline{\tau^2_{d]}} \tilde{e}^a\w \tilde{e}^b \w \tilde{e}^c \w \tilde{e}^d \Bigg) \ .\nn
\eea
We deduce from \eqref{zovz} and \eqref{oovo}
\beq
z \w \overline{z}\w \omega \w \overline{\omega} = -8\i e^{-4A}  \sum_{a,b,c=4,5,6} \re\left(\tau^1_{[1} \tau^2_{3]}\, \overline{\tau^1_{[a}} \overline{\tau^2_{b]}}\right) \tau^3_c \ \tilde{e}^1\w \tilde{e}^2 \w \tilde{e}^3 \w \tilde{e}^a\w \tilde{e}^b \w \tilde{e}^c \ . \label{resultcontradic}
\eeq
The structure compatibility conditions \eqref{compatjo} and \eqref{compatz} imply that the above form is proportional to the volume form and does not vanish
\beq
z \w \overline{z}\w \omega \w \overline{\omega} \neq 0 \ . \label{structurecond}
\eeq
We now show that this requirement does not hold, leading to a contradiction. From \eqref{result2}, we deduce the following, $\forall a,b=4,5,6$,
\beq
0=\tau^1_{\big[[3} \tau^2_{\phantom{\big[}b]} \tau^1_{a\big]} = \tau^1_{\big[3} \tau^2_{\phantom{\big[}b} \tau^1_{a\big]} =\frac{1}{3}\left(\tau^1_3 \tau^2_{[b} \tau^1_{a]} + \tau^1_a \tau^2_{[3} \tau^1_{b]} + \tau^1_b \tau^2_{[a} \tau^1_{3]} \right) =\frac{1}{3} \tau^1_3 \tau^2_{[b} \tau^1_{a]} \ .
\eeq
Thus, either $\tau^1_3=0$ or $\tau^2_{[b} \tau^1_{a]}=0$, so we can write
\beq
\tau^1_3\, \overline{\tau^2_{[b}} \overline{\tau^1_{a]}}=0 \ .
\eeq
We proceed similarly with $0=\tau^2_{\big[[3} \tau^1_{\phantom{\big[}b]} \tau^2_{a\big]}$ $\forall a,b=4,5,6$, and obtain eventually $\tau^2_3 \, \overline{\tau^1_{[b}} \overline{\tau^2_{a]}}=0$. From these two results, we deduce
\beq
\forall a,b=4,5,6,\quad \tau^1_{[1} \tau^2_{3]}\, \overline{\tau^1_{[a}} \overline{\tau^2_{b]}}=0 \ .\label{resultccl}
\eeq
This result implies that the combination \eqref{resultcontradic} vanishes, contradicting the non-vanishing of the form \eqref{structurecond}.

While the way to reach a contradiction is formal, the point can be understood intuitively as follows. Typically, $z, z^1, z^2$ should be independent (1,0)-forms with respect to an almost complex structure, thus spanning the six real dimensions. \eqref{result1} implies that the direction $\tilde{e}^3$ is not captured by $z$ and thus rather by $z^1, z^2$. If it were the case, meaning typically that $\tilde{e}^3$ is the real or imaginary of one of the $z^a$, one would get at least two terms in $\omega$ depending on $\tilde{e}^3$. However \eqref{result2} indicates that $\tau^1_{[3} \tau^2_{a]}$ is only non-zero for $a=1$, meaning that only one term of $\omega$ contains $\tilde{e}^3$, as in \eqref{ome}. This apparent contradiction is a sign of an independency or basis problem among the forms, which typically translates into the volume form vanishing. This is indeed what we showed.

As a side remark, note that this proof can be extended to the manifolds (with $\tilde{f}^3{}_{12} \neq 0$)
\beq
\d \tilde{e}^1=\d \tilde{e}^2=0\ , \ \d \tilde{e}^3= - \tilde{f}^3{}_{12}\, \tilde{e}^1\w \tilde{e}^2 -\sum_{b,c\neq2} \frac{1}{2} \tilde{f}^3{}_{bc}\, \tilde{e}^b \w \tilde{e}^c\ , \ \d \tilde{e}^{a=4,5,6}= -\sum_{b,c\neq2} \frac{1}{2} \tilde{f}^a{}_{bc}\, \tilde{e}^b \w \tilde{e}^c \ .\nn
\eeq
However, this is not useful here since the $O_4$ sets to zero (see \eqref{manifold}) the above additional structure constants with respect to \eqref{niltorus}.\\

We conclude on the following no-go: there is no supersymmetric Minkowski vacuum with an orthogonal SU(2) structure on the manifold $N_3 \times T^3$ described by \eqref{niltorus}, and with a space-time filling $O_4$-plane wrapping the internal $\tilde{e}^2$ direction. Since orthogonal SU(2) is the only constant SU(3)$\times$SU(3) structure allowed by the $O_4$, and the non-constant ones are very unlikely to exist, it is highly probable that there is no supersymmetric Minkowski vacuum {\it at all} on $N_3 \times T^3$ with an $O_4$ along $\tilde{e}^2$. This will be confirmed in section \ref{sec:nogo}, even without supersymmetry. The crucial point appears already to be the direction wrapped by the source: letting it wrap rather $\tilde{e}^3$ gives the vacuum presented above; we will come back to this.

\subsection{General analysis}\label{sec:gensearchsusy}

We have just proven the absence of an appropriate supersymmetric Minkowski vacuum on $N_3 \times T^3$; we now turn to the more general manifold \eqref{manifold}. Proceeding similarly, it is difficult to get to the same conclusion. This is illustrated by the manifold considered in section 3.2 of \cite{Andriot:2015sia} which is a particular case of \eqref{manifold}: as explained at the end of that section, the whole set of supersymmetry conditions can be solved for that example, and one is only left in the end with studying the BI. As discussed in section \ref{sec:condsusy}, the BI can be difficult to satisfy, and in that example, a problem occurs only in that of $F_2$. Conducting the same analysis in the more general case of \eqref{manifold} looks then too involved. Still, we can extract some useful information from the supersymmetry conditions.

The two conditions \eqref{susy1} and \eqref{susy3} simply lead to constraints on the coefficients. Let us treat the other conditions: using the algebra \eqref{manifold} for \eqref{susy2}, the form of $j$ \eqref{structforms} and the algebra for \eqref{susy4}, the compatibility condition \eqref{compatjo} and \eqref{susy1} and the dependence of $A$ for \eqref{susy5}, we obtain respectively
\beq
 H\w \omega = 0\ ,\quad  H\w \re(z) = 0\ ,\quad  H\w \im(z)\w j =0 \ .
\eeq
From this and \eqref{susy9}, we deduce that
\beq
F_0=0 \ .
\eeq
$F_2$ is given by the complicated \eqref{susy8}. Finally, $F_4$ is given by \eqref{susy7}
\beq
g_s * F_4= - e^{4A} \d (e^{-4A}) \w \tilde{e}^2 \ .\label{F4}
\eeq
In short, the supersymmetry conditions fix some of the free coefficients, and the flux $F_4$. The only other non-zero fluxes are $H$ and $F_2$, and they can only be constrained further by studying their BI, as explained above. It is difficult to do so in the general case \eqref{manifold}. Let us now rather focus on $F_4$.

The expression \eqref{F4} for $F_4$ is very particular and we will comment on it; let us first obtain its BI and study its consequences. The structure forms $z, \omega, j$ \eqref{structforms} and the almost complex structure allow to build a metric in the basis $(\tilde{e}^2, \tilde{e}^a_{\bot})$ as described in \cite{Andriot:2015sia}. Since only $\im z$ depends on $\tilde{e}^2$, and depends on $\tilde{e}^2$ only, this metric is block diagonal: it is given in \eqref{internalmetric}. We can then relate the six-dimensional internal Hodge star $*$ to the five-dimensional one along the orthogonal directions $*_{\bot}$ as follows
\beq
g_s F_4= - e^{4A} * (\d (e^{-4A}) \w \tilde{e}^2)= - e^{3A} *_{\bot} \d (e^{-4A}) = - \tilde{*}_{\bot} \d (e^{-4A}) \ ,
\eeq
where in the last equality, we used the tilde (smeared) metric; it results in the warp factor dependence dropping out. The exterior derivative is then given by
\beq
g_s \d F_4= - \d (\tilde{*}_{\bot} \d (e^{-4A})) =  - \tilde{\Delta}_{\bot}(e^{-4A})\ \tilde{*}_{\bot} 1  \ ,
\eeq
where our conventions give in general $\Delta f= * \d * \d f$ for a function $f$.

In the case $H=F_2=0$, the BI for $F_4$ is only given by $\d F_4$ that equals the source contributions. By integrating it on the transverse directions, or taking the smeared limit where $e^A$ is constant, we see that the source contributions from $O_4$ and $D_4$ should cancel each other. Looking at the Einstein equation on the internal directions, we deduce the Ricci tensor vanishes in the smeared limit: indeed the flux $F_4$ vanishes, the sources cancel each other, and the derivatives of dilaton vanish. This implies that the underlying solvmanifold is Ricci flat; this is precisely the reasoning followed in \cite{Andriot:2015sia}. However, only three solvable algebras allow for solvmanifolds with Ricci flat metrics, and none of them has the form \eqref{manifold}: since $\tilde{f}^3{}_{12} \neq 0$, these algebras would require $\tilde{f}^1{}_{32} \neq 0$ or $\tilde{f}^2{}_{31} \neq 0$, which is not the case here. We conclude that there is no appropriate supersymmetric Minkowski vacuum on the general manifold \eqref{manifold} in the case $H=F_2=0$. If those fluxes are non-zero, they would typically be related to each other, and cancel the source contributions in the BI as in the Einstein equation, leading to the same reasoning and conclusion. We refrain from showing the latter more formally here, and will prove the complete absence of appropriate Minkowski vacua using different tools in the next section.

\subsubsection*{Interpretation}

We can provide already an explanation for the absence of an appropriate supersymmetric Minkowski vacuum; the relevant ideas were discussed in \cite{Andriot:2015sia}. As mentioned, the $F_4$ obtained \eqref{F4} is very unusual: it is given by a derivative of the warp factor, and thus vanishes in the smeared limit where $e^A$ is constant. This should be compared to the more standard solution given in section \ref{sec:N3T3} where the $F_4$ \eqref{F4standard} has a constant term, so does not vanish in the smeared limit. This difference is reflected in the BI, and whether the source contributions cancel each other or not in the smeared limit, a point we have just used in the reasoning above. As pointed out in section \ref{sec:N3T3}, this difference, namely the absence or presence of the constant terms, should be related to the internal directions wrapped by the sources. This relation was established in \cite{Andriot:2015sia} and we reproduce now the argument, for $H=0$. We denote with a tilde the polyform from which we have extracted all warp factor dependence: consider for instance for the $p$-form part that there is a power $n$ of warp factor: $\im(\Phi_{\mp})|_p= e^{nA} \im(\widetilde{\Phi}_{\mp})|_p$; consider also that $e^{\phi}=g_s e^{mA}$. We then get from the last supersymmetry condition in \eqref{SUSY}
\bea
\pm \frac{g_s}{8}\, *\lambda(F_{5-p}) &=  e^{-4A}\, \d \left(e^{(3-m+n)A}\im(\widetilde{\Phi}_{\mp})|_p \right)\\
&=  e^{-4A}\, \d \left(e^{(3-m+n)A} \right) \w \im(\widetilde{\Phi}_{\mp})|_p +  e^{(-1-m+n)A}\, \d \im(\widetilde{\Phi}_{\mp})|_p  \ .\nn
\eea
The constant terms in the flux now clearly come from the very last term, namely $\d \im(\widetilde{\Phi}_{\mp})|_p$. Because of the condition for the supersymmetry of the source \eqref{calib}, one can relate this to the directions wrapped:
\beq
\d\, P[\im(\widetilde{\Phi}_{\mp})|_p]= \frac{1}{8} \d \widetilde{{\rm vol}}_{p\, ||}\ .
\eeq
The presence of constant terms in the flux are thus related to whether the (smeared) volume wrapped by the source is closed or not. This is precisely the difference in the two cases considered in section \ref{sec:N3T3}: $\d \tilde{e}^3 \neq 0$, but in the appropriate case for us, $\d \tilde{e}^2 = 0$. In other words, because we are looking for sources with $\d \widetilde{{\rm vol}}_{||}= 0$, we are then lead to fluxes that vanish in the smeared limit. A priori, this is not a problem, but as argued in \cite{Andriot:2015sia} and above, this is typically realised on Ricci flat solvmanifolds. The embedding of the nilmanifold $N_3$ in $\mmm$ rather results in an underlying solvmanifold \eqref{manifold} which is not Ricci flat, hence the problem. This tension between the curvature of the manifold on one hand, and the direction wrapped by the sources reflected in the flux on the other hand, will be as well the main reason for the no-go theorem in section \ref{sec:nogo}, as mentioned in the Conclusion.

\section{General search for an appropriate vacuum}\label{sec:gensearch}

\subsection{Manipulations of the equations of motion}

In order to establish the no-go theorem in section \ref{sec:nogo}, we need a specific relation between the vacuum quantities. We derive here this relation by various manipulations of the equations of motion. We start with the (bosonic) type IIA supergravity action in string frame, from which the Einstein equation and dilaton equation of motion are derived
\beq
S=\frac{1}{2\kappa^2} \int \d^{10} x \sqrt{|g_{10}|}\ \left( e^{-2\phi} ({\cal R}_{10} +4 |\del{\phi}|^2  - \frac{1}{2} |H|^2 )
 - \frac{1}{2} (|F_0|^2 +|F_2|^2+ |F_4|^2)  \right)\ ,
\eeq
where $\ 2\kappa^2=(2\pi)^7 (\alpha^\prime)^4$, $\alpha^\prime=l_s^2$, and $|g_{10}|$ is the absolute value of the determinant of the ten-dimensional metric. We denote for the $p$-form $F_p$ in ten dimensions
\beq
F_p \w *_{10} F_p=\d^{10} x\ \sqrt{|g_{10}|} \ \frac{F_{M_1 \dots M_p} F^{M_1 \dots M_p}}{p!} = \d^{10} x\  \sqrt{|g_{10}|} \ |F_p|^2 \ .
\eeq
To the above should be added the topological Chern-Simons term, but it does not contribute to the Einstein nor the dilaton equation of motion. In addition, one should consider the action for the sources ($D_p$ or $O_p$); for the same reason we only need the DBI part and not the Wess-Zumino term
\beq
S_{DBI}=- c_p\,  T_p \int_{\Sigma_{p+1}} \d^{p+1}\xi \ e^{-\phi} \sqrt{|P[g_{10} + b] + \mathcal{F}|}  \ ,
\eeq
where the action is on the world-volume $\Sigma_{p+1}$ and $P[ \cdot]$ is the pull-back to it. The tension $T_p$ is given by $T_p^2=\frac{\pi}{\kappa^2} (4\pi^2 \alpha^\prime)^{3-p}$ and for a $D_p$, $c_p =1$ while for an $O_p$, $c_p =-2^{p-5}$ and $\mathcal{F}=0$. It is useful to bring this action to a ten-dimensional one. For the cases of interest here, there is no complication in the embedding of the sources in the ten-dimensional space-time, so the pull-back is only a projection on $\Sigma_{p+1}$; the sources will be space-time filling so $\Sigma_{p+1}$ is the product of the four-dimensional space-time and the internal subspace wrapped. It is then enough to complete to ten dimensions by a $\delta(y_{\bot})$ localizing in the internal directions transverse to the source
\beq
S_{DBI}=- c_p\, T_p \int \d^{10}x\ e^{-\phi} \sqrt{|P[g_{10} + b] + \mathcal{F}|} \ \delta(y_{\bot})  \ . \nn
\eeq
We can now derive the sources contributions to the equations of motion. We introduce the energy momentum tensor
\beq
\frac{1}{\sqrt{|g_{10}|}} \sum_{{\rm sources}} \frac{\delta S_{DBI}}{\delta g^{MN}}=- \frac{e^{-\phi}}{4 \kappa^2} T_{M N} \ ,
\eeq
where ${}_{M, N, ...}$ are ten-dimensional curved indices. We restrict from now on to $P[b] + \mathcal{F} = 0$. The DBI action is then essentially given by the determinant of the metric on the world-volume, so $T_{MN}$ is given by the standard result with a projector towards $\Sigma_{p+1}$, and we also deduce its trace
\bea
& T_{M N}= - \frac{2\kappa^2}{\sqrt{|g_{10}|}} \sum_{{\rm sources}} c_p\,  T_p\ P[g_{MN}] \sqrt{|P[g_{10}]|} \ \delta(y_{\bot}) \ ,\\
& T_{10}= g^{MN} T_{M N} = - \frac{2\kappa^2}{\sqrt{|g_{10}|}} \sum_{{\rm sources}} c_p\,  T_p\ (p+1) \sqrt{|P[g_{10}]|} \ \delta(y_{\bot}) \ .
\eea
One can then verify\footnote{For sources of different sizes $p$, the quantity $\frac{T_{10}}{p+1}$ should be taken as a notation, since the quotient should then be realised within the sum on sources. We however soon restrict to only one value of $p$.}
\beq
\frac{1}{\sqrt{|g_{10}|}} \sum_{{\rm sources}} \frac{\delta S_{DBI}}{\delta \phi}=- \frac{e^{- \phi}}{2 \kappa^2} \frac{T_{10}}{p+1} \ .
\eeq
The Einstein and the dilaton equation of motion are then\footnote{The dilaton terms in the first line of \eqref{Einst} might not be obvious to derive; we refer for instance to the footnote 30 of \cite{Andriot:2011iw} about them. We also recall the definition of the, say six-dimensional, Laplacian on a function $\varphi$: $\Delta \varphi = g^{mn} \nabla_m \del_n \varphi = \frac{1}{\sqrt{|g|}} \del_m (\sqrt{|g|} g^{mn} \del_n \varphi )$. Finally, note that the democratic formalism is often used to derive supergravity equations of motion, instead of the above action; this is actually required for the RR sector, but we do not need the corresponding equations here.}
\bea
{\cal R}_{MN}-\frac{g_{MN}}{2} {\cal R}_{10} & = \frac{e^{\phi}}{2}T_{MN} -2\nabla_M \del_N{\phi} +2 g_{MN} (\Delta \phi - 2 |\del \phi|^2 ) \label{Einst} \\
 & +\frac{1}{4} H_{MPQ}H_N^{\ \ PQ}+\frac{e^{2\phi}}{2}\left(F_{2\ MP}F_{2\ N}^{\ \ \ \ P} +\frac{1}{3!} F_{4\ MPQR}F_{4\ N}^{\ \ \ \ PQR} \right) \nn\\
& -\frac{g_{MN}}{2} \left(-4|\del \phi|^2+ \frac{1}{2} |H|^2 + \frac{e^{2\phi}}{2}(|F_0|^2 + |F_2|^2 + |F_4|^2 )\right)   \ , \nn\\
&\nn\\
2 {\cal R}_{10} + 8(\Delta \phi - |\del \phi|^2 )& -|H|^2 = -e^{\phi} \frac{{T}_{10}}{p+1} \ .\label{dileom}
\eea
Let us point-out that the RR fluxes appearing here are, in our case, the same as those introduced in section \ref{sec:susy}, as explained in footnote \ref{foot:demo}.

We now trace the above Einstein equation in ten and in four dimensions; for the latter we recall that the fluxes are purely internal. We denote
\beq
{\cal R}_{10}= g^{MN} {\cal R}_{MN} \ ,\ {\cal R}_4= g^{\mu\nu} {\cal R}_{\mu\nu} \ ,\ {\cal R}_6= g^{mn} {\cal R}_{mn}={\cal R}_{10} - {\cal R}_{4} \ , \ (\nabla\del \phi)_4= g^{\mu\nu}\nabla_{\mu}\del_{\nu} \phi \ .
\eeq
The dilaton is purely internal so the quantity $(\nabla\del \phi)_4$ would naively vanish. We are however using the ten-dimensional quantities (here the connection in $\nabla_M$ in particular) and look at four-dimensional indices; this is different from pure four-dimensional quantities. The same holds for the seemingly Ricci scalars just defined. We get
\bea
& 4 {\cal R}_{10}  + \frac{e^{\phi}}{2} {T}_{10} -20 |\del \phi|^2 + 18 \Delta \phi - |H|^2 - \frac{e^{2\phi}}{2}(5|F_0|^2 + 3|F_2|^2 + |F_4|^2 ) = 0\ ,\\
& {\cal R}_4 - 2{\cal R}_{10} - \frac{2 e^{\phi}}{p+1} {T}_{10} +2 (\nabla\del \phi)_4 + 8 |\del \phi|^2 - 8 \Delta \phi + |H|^2 + e^{2\phi} (|F_0|^2 + |F_2|^2 + |F_4|^2 ) = 0 \ ,
\eea
where we used
\beq
g^{\mu\nu} T_{\mu\nu} = \frac{4}{p+1} {T}_{10} \ .
\eeq
The supersymmetry conditions and the BI would allow to relate $|F_0|^2 + |F_2|^2 + |F_4|^2$ to $T_{10}$ following (A.15) of \cite{Andriot:2015sia}; we however want to be more general here so do not use such a relation. Instead, we now use the dilaton equation of motion to eliminate ${T}_{10}$ in respectively the ten- and four-dimensional traces.\footnote{Combining the dilaton equation of motion and the four-dimensional trace, one obtains
\beq
{\cal R}_4 = e^{\phi} \frac{{T}_{10}}{p+1} - e^{2\phi} (|F_0|^2 + |F_2|^2 + |F_4|^2 ) - 2 (\nabla\del \phi)_4 \ .
\eeq
In the smeared limit, $\phi$ is constant and ${\cal R}_4$ is proportional to the cosmological constant. Getting a de Sitter vacuum then requires not only the orientifolds, but also that their contribution is bigger than that of the $D_p$-branes, so that $\frac{{T}_{10}}{p+1} > 0$. To study the non-smeared case, one should first reconstruct $\tilde{\Delta} e^{-4A}$ and consider integrals. I thank T. Van Riet for related discussions.} We also restrict from now on to a set of sources with only one fixed value of $p$: $(p+1)$ can then be extracted of the sum on sources. We get
\bea
& (p-3)\left( -2  {\cal R}_{10} + 8 |\del \phi|^2 - 8 \Delta \phi + |H|^2 \right) \label{10dtracesansT10}\\
& -8 |\del \phi|^2 + 4 \Delta \phi + 2 |H|^2 - e^{2\phi}(5|F_0|^2 + 3|F_2|^2 + |F_4|^2 ) = 0  \ ,\nn \\
& 3 {\cal R}_4 = - 2{\cal R}_{6} -2 (\nabla\del \phi)_4 + 8 |\del \phi|^2 - 8 \Delta \phi + |H|^2 - e^{2\phi} (|F_0|^2 + |F_2|^2 + |F_4|^2 ) \ ,\label{4dtracesansT10}
\eea
where we wrote \eqref{10dtracesansT10} in a particular form to combine it with \eqref{4dtracesansT10}. We now multiply \eqref{4dtracesansT10} by $(p-3)$ and insert \eqref{10dtracesansT10} (note that $p-3 \neq 0$ as we are in IIA). This gives
\bea
(p-3) {\cal R}_4 & = 2 e^{2\phi} \Delta e^{-2\phi} -2 (p-3)(\nabla\del \phi)_4 \label{4dtracefinal} \\
 & - 2 |H|^2 + e^{2\phi} \left( (8-p) |F_0|^2 + (6-p) |F_2|^2 + (4-p) |F_4|^2 \right) \ , \nn
\eea
where we used $-2 |\del \phi|^2 + \Delta \phi= -\frac{1}{2} e^{2\phi} \Delta e^{-2\phi}$. For future purposes, we now simply sum \eqref{10dtracesansT10} to the right-hand side of \eqref{4dtracesansT10}, make use of $(-2\times)$\eqref{4dtracefinal}, and write the resulting equation as follows
\bea
2(p-2) {\cal R}_6 & = -(4p-9) {\cal R}_4 + 2 e^{2\phi} (|F_0|^2 - |F_4|^2) \label{final}\\
   & -2 (2p-5)(\nabla\del \phi)_4  + 8 (p-1) |\del \phi|^2 -4 (2p-3) \Delta \phi  \nn\\
 & +(4-p) \left(- |H|^2 + 2 e^{2\phi} \left( |F_0|^2 + |F_2|^2 + |F_4|^2 \right)\right) \ . \nn
\eea

\subsubsection*{Comment and check}

Equation \eqref{4dtracefinal} has an interesting interpretation: for $D_p$ or $O_p$ sourcing one of the RR-fluxes $F_q$, the coefficient in this equation in front of $F_q$ precisely vanishes. The dilaton terms would typically compensate the warp factor terms coming from ${\cal R}_4$. This leaves essentially the smeared ${\cal R}_4$ to be given by the other fluxes, which is an interesting relation.

This relation can be tested for instance for $p=4$ on Minkowski, where one would typically have $F_0=0$ while $H$ and $F_2$ compensate each other, if not vanishing: see the example in section \ref{sec:N3T3}. An extreme subcase is the $D_4$ background: let us check this formula for that example. There, the ten-dimensional metric is diagonal and only given by the warp factor. The brane is along the five first directions and orthogonal to the last five, and we recall the warp factor only depends on the orthogonal directions
\beq
g= {\rm diag}(- e^{2A}, e^{2A}, \dots, e^{2A}, e^{-2A}, \dots, e^{-2A} ) \ .
\eeq
Denoting as above by a tilde the metric without warp factor, i.e. the smeared version, we have here $\tilde{g}_{MN}= \eta_{MN}$. The dilaton is  $e^{\phi}=g_s e^A$ and the only non-zero flux is $F_4$, its components being proportional to $\del e^{-4A}$. The smeared Laplacian is simply given by $\tilde{\Delta}= \eta^{MN} \del_M \del_N$, and one has $\tilde{\Delta} e^{-4A}= - 2\kappa^2 c_p\,  T_p\, g_s \delta(y_{\bot}) $. We get
\bea
& \begin{cases} {}_{M=N= ||}:\ {\cal R}_{MN}= -\frac{1}{2}  \eta_{MN} e^{2A} \tilde{\Delta} e^{2A}\\
{}_{M=N= \bot}:\ {\cal R}_{MN}= \frac{e^{-4A}}{2} \left( -4  \del_M e^{2A} \del_N e^{2A} +  \eta_{MN} e^{2A} \tilde{\Delta} e^{2A} - 2 e^{2A} \del_M \del_N e^{2A} \right) \end{cases} ,\\
& {\cal R}_{10}= -2 e^{-2A} (\widetilde{\del e^{2A}})^2 - \tilde{\Delta} e^{2A} \ ,\ {\cal R}_4= -2 \tilde{\Delta} e^{2A} \ .
\eea
In addition, $\nabla_M V_N = \del_M V_N - \Gamma_{MN}^P V_P$ and
\beq
{\rm For}\ {}_{M,N= ||}, {}_{P= \bot},\ \Gamma_{MN}^P= -\frac{1}{2} \eta^{PQ} \eta_{MN} e^{2A} \del_Q e^{2A} \Rightarrow (\nabla\del \phi)_4 = e^{-2A} (\widetilde{\del e^{2A}})^2 \ .\label{truc}
\eeq
It is now straightforward to check the above relations, in particular \eqref{4dtracefinal} where only the first row is non-zero: there one has by definition $e^{2\phi} \Delta e^{-2\phi}= e^{2A} \eta^{MN} \del_M (e^{2A} \del_N e^{-2A}) $, so the relation is verified.

\subsection{No-go theorem}\label{sec:nogo}

In section \ref{sec:strategy}, we presented our strategy to embed the  monodromy mechanism into a concrete compactification setting: one should first find an appropriate (static) vacuum, corresponding to the minimum of the potential. In this section, using the tools developed, we prove a no-go theorem against finding such a vacuum. We first list all necessary assumptions, then give the theorem, and finally prove it.

\subsubsection*{Assumptions}

Let us first summarize the assumptions made so far.\\
- We consider ten-dimensional massive type IIA supergravity with $D_p$-branes and orientifold $O_p$-planes, and look for a vacuum. We do not consider any other ingredient such as $\NS5$-branes or $\KK$-monopoles,\footnote{\label{foot:KK}It is worth noticing that the four-dimensional approach of \cite{Gur-Ari:2013sba}, presented at the end of the Introduction, included those two ingredients and still reached conclusions similar to ours.} non-geometric contributions (exotic branes, fluxes...), fermionic contributions such as condensates, $\alpha'$ or $g_s$ or non-perturbative contributions, etc.\\
- The $D_p$ and $O_p$ are space-time filling, and have a pull-back given simply by a projection, i.e. there is no complication in their embedding. We further restrict to $P[b] + \mathcal{F} = 0$, and finally, to sources of only one size $p$.\\
- Having in mind the compactification setting, the vacuum looked for is on a ten-dimensional space-time given by a warped product with metric \eqref{10dmetric}. The fluxes $H, F_0, F_2, F_4$ are purely internal (in particular $F_4$).

These standard assumptions lead us to derive the above relations using the equations of motion, in particular \eqref{final} that we will use below. We now list further assumptions that are more specific to the vacuum we are looking for.\\
- The size of the sources is fixed to $p=4$. As explained in section \ref{sec:strategy}, there are $D_4$ but we also require $O_4$. Both should wrap the internal direction along $\tilde{e}^2$ because of the inflation mechanism. We take the standard corresponding dilaton $e^{\phi}=g_s e^A$, and $A$ depends on the transverse directions.\\
- The underlying manifold is a twisted torus (solvmanifold), and the compact manifold $\mmm$ only differs by warp factor rescaling. Given the sources, the internal metric is \eqref{internalmetric}; this implicitly means that no localized sources along other directions are considered, otherwise the warp factor dependence would differ. Note that vacua with intersecting sources are very difficult to construct without smearing. The solvmanifold should include $N_3$ accordingly to the mechanism: this together with the $O_4$ projection restricts the allowed solvmanifolds to \eqref{manifold}.\\
- The sources are supersymmetric and calibrated: their energy is then minimized, in other words the open string degrees of freedom are stabilized in a vacuum. This is actually a standard assumption, and going beyond would be difficult. For de Sitter vacua, supersymmetry is typically rather broken either by the bulk fields, or by non-mutually BPS branes, but each of them would still be separately BPS (for non-BPS branes, see e.g. \cite{Bergshoeff:2000dq}). Having sources as we require can also be viewed as necessary to get a supersymmetric theory in four dimensions as supergravity: supersymmetry is then spontaneously broken in the vacuum rather than explicitly in the four-dimensional theory.\footnote{An attempt to go beyond this standard requirement was made in \cite{Andriot:2011uh}: there, different configurations of sources were considered, where their energy is minimized, but they break supersymmetry by wrapping a subspace in an unusual manner. A characterisation of those was proposed, but such a situation is generally not well-understood.} This assumption on the sources has to be implemented differently according to the external space-time, so we now turn to this.
\begin{itemize}
\item For Minkowski and de Sitter four-dimensional space-time: for the former, the supersymmetry condition on the source was given in \eqref{calib}, and the calibration was then given by the last supersymmetry condition of \eqref{SUSY}. These requirements need an SU(3)$\times$SU(3) structure, but are weaker than asking for a supersymmetric vacuum. For this reason and as argued above, we make the same requirements for de Sitter. This results in fixing completely the sourced flux $F_4$ as we now show. The sources are along $\tilde{e}^2$ so one has ${\rm vol}_{||} =e^A \tilde{e}^2$, giving from \eqref{calib} the pull-back (here projection towards $\tilde{e}^2$) of the one-form $\im(\Phi_-)|_1$. In addition, the orientifold projection \eqref{O4projPhi} indicates that $\im(\Phi_-)|_1$ should be even under the involution: this implies that this one-form is along the source. We deduce that $\im(\Phi_-)|_1$ is only given by the volume form and nothing more, i.e. $\im(\Phi_-)|_1 = P[\im(\Phi_-)]$. We then deduce $F_4$ from the calibration, i.e. the last supersymmetry condition of \eqref{SUSY}
    \beq
    \hspace{-0.1in} g_s \frac{e^{4A}}{8}  * F_4 = \d (e^{2A}\im(\Phi_-)|_1) = \d (e^{2A}P[\im(\Phi_-)])= \frac{1}{8}\d (e^{3A} {\rm vol}_{||}) = \frac{1}{8}\d (e^{4A} \tilde{e}^2) \ ,
    \eeq
    without knowing the details of $\Phi_-$. The manifold \eqref{manifold} imposes $\d \tilde{e}^2 = 0$, but as discussed at the end of section \ref{sec:gensearchsusy}, this is also an important property of the sources. We get eventually
    \beq
    F_4= g_s^{-1} e^{-4A} * ( \d (e^{4A}) \w \tilde{e}^2) \ .\label{F4final}
    \eeq

\item For anti-de Sitter four-dimensional space-time: the sources supersymmetry and calibration conditions have been worked-out in \cite{Koerber:2007jb} and work somehow analogously: the condition \eqref{calib} should still be imposed and one of the anti-de Sitter supersymmetry conditions corresponds to the minimization of the energy. This last condition is corrected with respect to the Minkowski one by a term depending on the cosmological constant $\Lambda$. We get in our conventions in type IIA
    \beq
    \frac{e^{4A}}{8}  * \lambda( F) = (\d -H\w) (e^{3A-\phi}\im(\Phi_-)) - 3 e^{2A-\phi}\im(\ov{\mu}\Phi_+) \ ,
    \eeq
    where $3|\mu|^2=-\Lambda$. The interpretation of the additional term is however subtle: it can be viewed as a boundary term, related to the boundary of anti-de Sitter \cite{Koerber:2007jb}.\footnote{Another subtlety is the following: anti-de Sitter supersymmetric vacua with constant SU(3)$\times$SU(3) structure are constrained: in type IIA, only SU(3) structure is admitted. But the $O_4$-plane projection is only compatible with orthogonal SU(2) structure vacua. This opposition is reminiscent of the fact that orientifolds are not always required for anti-de Sitter vacua, even though we consider one here (see section \ref{sec:strategy}). This contradiction is not a problem here as we do not ask for a supersymmetric vacuum, but it is worth noting it.} Projecting the above condition on the two-form part and using previous results, we get eventually
    \beq
    g_s F_4 = e^{-4A} * (\d (e^{4A}) \w \tilde{e}^2) - 3 e^{-2A} *\im(-\i e^{\i \theta_+} \ov{\mu}\omega) \ . \label{F4finalads}
    \eeq
\end{itemize}
Using this list of reasonable assumptions, we obtained the relation \eqref{final}, and will be able to compute most of its terms. This will lead to the proof of the no-go theorem that we now state.

\subsubsection*{No-go theorem}

The list of assumptions just detailed characterises a specific type of vacua, that we argued would be appropriate to embed the monodromy inflation mechanism of \cite{Silverstein:2008sg} in a concrete compactification setting. The no-go theorem on such a vacuum is now given as follows:\\

\noindent- There is no such vacuum with four-dimensional anti-de Sitter or Minkowski space-time.\\
- Such a vacuum with four-dimensional de Sitter space-time is not excluded, but there is a lower bound on the value of its cosmological constant: this bound is too high for phenomenology.

\subsubsection*{Proof}

To prove this no-go theorem, we use the above assumptions and the relation \eqref{final}: for $p=4$, that relation becomes
\beq
4 {\cal R}_6  = -7 {\cal R}_4 + 2 e^{2\phi} (|F_0|^2 - |F_4|^2)  -6(\nabla\del \phi)_4  + 24 |\del \phi|^2 - 20 \Delta \phi  \ . \label{rel}
\eeq
Let us now compute the various terms. We recall the ten-dimensional metric given by \eqref{10dmetric}
\beq
\d s^2= e^{2A(y)} \tilde{g}_{\mu\nu} (x) \d x^\mu \d x^\nu + g_{mn} (y) \d y^m \d y^n \ ,
\eeq
and we denote as always with a tilde the smeared (unwarped) quantities: $\tilde{g}_{MN}=g_{MN}|_{A=0}$. The six-dimensional metric $g_{mn}$ does not need to be specified for now, except for its determinant that verifies $\sqrt{|g_6|}=e^{-4A}\sqrt{|\tilde{g}_6|} $ because the $D_4$ and $O_4$ wrap one internal direction. Another useful property is that the warp factor depends only on the transverse directions, implying $g^{mp}\del_p e^{2A}= e^{2A} \tilde{g}^{mp}\del_p e^{2A}$. One also has
\beq
\tilde{\Delta}_{10} e^{2A}=\frac{1}{\sqrt{|\tilde{g}_{10}|}} \del_M \left(\sqrt{|\tilde{g}_{10}|} \tilde{g}^{MN} \del_N e^{2A} \right)=\frac{1}{\sqrt{|\tilde{g}_{6}|}} \del_m \left(\sqrt{|\tilde{g}_{6}|} \tilde{g}^{mn} \del_n e^{2A} \right)=\tilde{\Delta}_{6} e^{2A} \equiv \tilde{\Delta} e^{2A} \ .\nn
\eeq
Using these properties and standard expressions for the Ricci tensor with Levi-Civita connection (see e.g. \cite{Andriot:2013xca}), we get
\bea
& {\cal R}_{\mu\nu}=\tilde{{\cal R}}_{\mu\nu} - \frac{1}{2} \tilde{g}_{\mu\nu} \Delta_6 e^{2A} - \frac{1}{2} \tilde{g}_{\mu\nu} e^{-2A} (\del e^{2A})^2 =\tilde{{\cal R}}_{\mu\nu} - \frac{1}{2} \tilde{g}_{\mu\nu} e^{2A} \tilde{\Delta} e^{2A} \ ,\\
& {\cal R}_4= e^{-2A} \tilde{{\cal R}}_4  -2 \tilde{\Delta} e^{2A} \ .
\eea
Further, denoting as before ${\cal R}_{mn}={\cal R}_{MN=mn}$, and $R_{mn}$ the six-dimensional Ricci tensor constructed directly from $g_{mn}$, we obtain
\bea
& {\cal R}_{mn} = R_{mn} - 2 \nabla_n (e^{-2A} \del_m e^{2A}) - e^{-4A} \del_m e^{2A} \del_n e^{2A}\ ,\\
& {\cal R}_6= R_6 - 2 \tilde{\Delta} e^{2A} + 3 e^{-2A} (\widetilde{\del e^{2A}})^2 \ .
\eea
Computing the purely internal Ricci scalar $R_6$ of $\mmm$ is more involved, and now requires to specify the internal metric: it was given in \eqref{internalmetric}. We can use the expression of the Ricci scalar \eqref{Ricciflat} in terms of the vielbeins $e^a{}_m$ and related structure constants $f^{a}{}_{bc}$. From it, we can get an expression in terms of $\tilde{e}^a{}_m$ and $\tilde{f}^{a}{}_{bc}$, where we have extracted the warp factor to reach the underlying solvmanifold. We start by computing
\bea
& f^{2}{}_{b_{\bot}2}= e^A \tilde{f}^{2}{}_{b_{\bot}2} - \del_{\tilde{b}_{\bot}} e^A \ ,\quad f^{2}{}_{b_{\bot} c_{\bot}}= e^{3A} \tilde{f}^{2}{}_{b_{\bot} c_{\bot}} \ ,\nn\\
& f^{a_{\bot}}{}_{2 c_{\bot}}= e^{-A} \tilde{f}^{a_{\bot}}{}_{2 c_{\bot}} \ ,\quad f^{a_{\bot}}{}_{b_{\bot} c_{\bot}}= e^{A} \tilde{f}^{a_{\bot}}{}_{b_{\bot} c_{\bot}} +2 \delta^{\tilde{a}_{\bot}}_{[\tilde{c}_{\bot}} \del_{\tilde{b}_{\bot}]} e^A \ . \nn
\eea
We can then derive from \eqref{Ricciflat} a general formula for $R_6$. However, this is not necessary here: we can use the information on the allowed manifolds, namely \eqref{manifold}. This gives
\beq
\tilde{f}^{2}{}_{b_{\bot}2}= \tilde{f}^{2}{}_{b_{\bot} c_{\bot}} = \tilde{f}^{a_{\bot}}{}_{b_{\bot} c_{\bot}} = 0 \ .
\eeq
In addition, the unimodularity criterion $\tilde{f}^{a}{}_{ba}=0$, needed for compactness, gives $\tilde{f}^{a_{\bot}}{}_{2 a_{\bot}}=0$. From \eqref{Ricciflat}, we are left in the end with
\beq
R_6= e^{-2A} \tilde{{\cal R}}_6 + 6 \eta^{a_{\bot} b_{\bot}} e^A \del_{\tilde{a}_{\bot}} \del_{\tilde{b}_{\bot}} e^{A} -14 (\widetilde{\del e^{A}})^2 \ ,
\eeq
where $\tilde{{\cal R}}_6$ is the pure underlying solvmanifold Ricci scalar, given by
\beq
\tilde{{\cal R}}_6 = -\frac{1}{2} \tilde{f}^{a_{\bot}}{}_{2 b_{\bot}} \tilde{f}^{b_{\bot}}{}_{2 a_{\bot}} -\frac{1}{2} \eta_{a_{\bot} d_{\bot}} \eta^{c_{\bot} g_{\bot}} \tilde{f}^{a_{\bot}}{}_{2 c_{\bot}} \tilde{f}^{d_{\bot}}{}_{2 g_{\bot}} \ . \label{finalinternalR6}
\eeq
The $\del_{\tilde{a}_{\bot}} \del_{\tilde{b}_{\bot}}$ can be completed to a covariant derivative term, but the spin connection term, multiplied by the $\eta^{a_{\bot} b_{\bot}}$, is related to $\tilde{f}^{a_{\bot}}{}_{a_{\bot} b_{\bot}}$ that vanishes. We get eventually
\bea
& R_6= e^{-2A} \tilde{{\cal R}}_6 + 3 \tilde{\Delta} e^{2A} -5 e^{-2A}(\widetilde{\del e^{2A}})^2 \ ,\\
& {\cal R}_6 = e^{-2A} \tilde{{\cal R}}_6 + \tilde{\Delta} e^{2A} -2 e^{-2A}(\widetilde{\del e^{2A}})^2 \ .
\eea
Finally, using the expression for the dilaton and $\sqrt{|g_{10}|}=\sqrt{|\tilde{g}_{10}|}$ (see also \eqref{truc}), we get
\beq
(\nabla\del \phi)_4 = e^{-2A} (\widetilde{\del e^{2A}})^2 \ ,\ |\del \phi|^2 = \frac{1}{4} e^{-2A} (\widetilde{\del e^{2A}})^2 \ ,\ \Delta \phi= \frac{1}{2} \tilde{\Delta} e^{2A} \ .
\eeq
The initial relation \eqref{rel} now boils down to
\beq
4 e^{-2A} \tilde{{\cal R}}_6  = -7 e^{-2A} \tilde{{\cal R}}_4  + 2 e^{2\phi} (|F_0|^2 - |F_4|^2) + 8 e^{-2A}(\widetilde{\del e^{2A}})^2   \ . \label{rel2}
\eeq
We are left with computing $|F_4|^2$ for which we use the assumption on the supersymmetric and calibrated sources stated above. For Minkowski and de Sitter, we use \eqref{F4final} that gives
\beq
|F_4|^2=g_s^{-2} e^{-8A} |\d (e^{4A}) \w \tilde{e}^2|^2 = g_s^{-2} e^{-10A} (\del e^{4A})^2 = 4 g_s^{-2} e^{-4A} (\widetilde{\del e^{2A}})^2 \ ,
\eeq
while for anti-de Sitter, we use \eqref{F4finalads}: the additional term there is orthogonal to the first one (the internal metric is block diagonal) so we simply get a second contribution to the square
\beq
|F_4|^2= 4 g_s^{-2} e^{-4A} (\widetilde{\del e^{2A}})^2 + 6 g_s^{-2} e^{-4A} (-\Lambda)  \ .
\eeq
As expected, we see that in both cases, all dilaton and warp factor terms in \eqref{rel2} get canceled when including the $F_4$.

Having computed the terms of \eqref{rel} or \eqref{rel2}, we now introduce two last ingredients. We start with the cosmological constant: to get the latter, one should go to the four-dimensional Einstein frame; let us give details about it. Consider four-dimensional fluctuations of the internal volume and the dilaton with respect to vacua quantities, captured by the following scalar fields $\rho$, $\tilde{\phi}$ and $\sigma$
\beq
g_{mn}=\rho\, g_{mn}|_{{\rm vacuum}}\ ,\ \int_{\mmm} \d^6 y \sqrt{|g_6|}= v_6\, \rho^3 \ , \ e^{\phi}=g_s e^A\, e^{\tilde{\phi}}\ ,\ \sigma=\rho^{\frac{3}{2}} e^{-\tilde{\phi}} \ .
\eeq
In the vacuum, $\rho=\sigma=1$. To find the rescaling towards the four-dimensional Einstein frame with metric $g_{E\mu\nu}$, we then develop the following term of the ten-dimensional action
\beq
\int \d^{10}x \sqrt{|g_{10}|} e^{-2\phi} {\cal R}_{10} = g_s^{-2} v_6 \int_4 \d^4 x \sqrt{|\tilde{g}_{4}|} \sigma^2 \tilde{{\cal R}}_4 + \dots \ ,
\eeq
where the warp factor had to be extracted to end with a four-dimensional integral.\footnote{More generally, for parallel sources of dimension $p$, the typical vacuum dilaton is $e^{\phi}=g_s e^{(p-3)A}$ leading to a $e^{-2(p-4)A}$ in the integral defining $v_6$.} The rescaling bringing the above to the standard Einstein-Hilbert term is then $g_{E\mu\nu}=\sigma^2 \tilde{g}_{\mu\nu}$. In the vacuum, the cosmological constant is given by $R_{E4}=4\Lambda$, and then simply here $\tilde{{\cal R}}_4= 4\Lambda$; we refer to \cite{Andriot:2010ju} for more details.\footnote{In a four-dimensional approach, one has a scalar potential for $\rho$ and $\sigma$ as e.g. in section 3.4.1 of \cite{Andriot:2010ju}. In that case, identities derived here from the ten-dimensional equations of motion correspond to the potential being extremised, and $\Lambda$ is related to the vacuum value of the potential in Einstein frame. This potential depends on $-\sigma^{-2} \rho^{-1} \tilde{{\cal R}}_6$. For a large volume, i.e. a large $\rho$, this quantity is small, thus seemingly lowering the bound appearing in \eqref{finalL} corresponding here to a bound on the minimum of the potential. Nevertheless, the argument presented at the end of this section still holds (actually the $\rho^{-1}$ can be viewed as coming from the $(\tilde{r}_0)^{-2}$): the volume cannot be sent to infinity simply because of phenomenological bounds, i.e. the volume of the internal manifold is constrained by current observations and can certainly not reach cosmological length scales.} Secondly, an important element is the curvature of solvmanifolds. As argued in section \ref{sec:gensearchsusy}, the allowed solvmanifolds \eqref{manifold} are not Ricci flat and are thus negatively curved. This can be seen because the structure constants are such that the contributions to \eqref{finalinternalR6} either cancel each other, or are negative (see e.g. \cite{Andriot:2015sia}). In particular, given \eqref{manifold}, we get here
\beq
\tilde{{\cal R}}_6 \leq \tilde{{\cal R}}_{6_0} = -\frac{1}{2} (\tilde{f}^3{}_{12})^2 < 0 \ .
\eeq
We now conclude using \eqref{rel2}, the expressions for $|F_4|^2$, and the last two ingredients.
\begin{itemize}
\item For anti-de Sitter, one obtains
\beq
e^{-2A} \tilde{{\cal R}}_6  = -4 e^{-2A}\Lambda + \frac{1}{2} e^{2\phi} |F_0|^2 \ ,
\eeq
from which we deduce
\beq
4 \Lambda \geq -\tilde{{\cal R}}_{6_0} > 0 \ .
\eeq
This is impossible since $\Lambda <0$. This excludes anti-de Sitter vacua.

\item For Minkowski or de Sitter, one gets
\beq
e^{-2A} \tilde{{\cal R}}_6  = -7 e^{-2A}\Lambda + \frac{1}{2} e^{2\phi} |F_0|^2 \ , \label{relfinal}
\eeq
from which we deduce
\beq
7 \Lambda \geq - \tilde{{\cal R}}_{6_0} > 0 \ . \label{finalL}
\eeq
This excludes Minkowski vacua, and gives a lower bound to $\Lambda$ for de Sitter vacua. Let us now discuss this bound.

\end{itemize}

The internal curvature $\tilde{{\cal R}}_{6_0}$ of the underlying solvmanifold depends on $\tilde{f}^3{}_{12}$, so it can be expressed in terms of the three radii along directions $1,2,3$, and an integer $M$ as in section \ref{sec:inflation}. But in first approximation, it is of order $- \tilde{{\cal R}}_{6_0} \sim \frac{1}{(\tilde{r}_0)^2}$, with an average radius $\tilde{r}_0$ of the underlying solvmanifold. Then, a first conclusion is that it is impossible for the bound \eqref{finalL} to hold phenomenologically: indeed, inverting \eqref{finalL}, it would mean that the cosmological distances should be smaller than the radius $\tilde{r}_0$, which makes no sense; equivalently, the cosmological constant energy should be bigger than an internal energy, which cannot be satisfied given the observations. One could however milden the claim, because one needs to be careful on what are the actual internal distances of $\mmm$: the warp factor is also involved, and it may create a hierarchy \cite{Giddings:2001yu}. This remains however difficult to realise here, because different directions scale differently with the warp factor, as can be seen in the internal metric \eqref{internalmetric}. Indeed, suppose that \eqref{finalL} is phenomenologically realised, $\tilde{r}_0$ then needs to be a cosmological length: this can be in principle accommodated by choosing $e^{A}$ such that, say, $e^{-A} \tilde{r}_0$ is a typical internal length. But another internal direction then has $e^{A} \tilde{r}_0$ for radius, which is even more cosmological, so this is not phenomenologically viable. The warp factor can thus not create the necessary hierarchy and we are left with the first conclusion: the bound \eqref{finalL} cannot be satisfied phenomenologically.

\section{Conclusion}

In this paper, we have been interested in the  monodromy inflation mechanism of \cite{Silverstein:2008sg}. This mechanism generates an inflaton potential compatible with the current observations. We have attempted to embed it in a concrete compactification of string theory, in order to have a complete model. After having introduced some necessary material on the internal geometry in section \ref{sec:geom} and summarized the mechanism in section \ref{sec:inflation}, we have presented in section \ref{sec:strategy} our strategy for realising this embedding. The mechanism involves a moving brane reproducing the inflaton rolling down its potential towards a vacuum. We considered the particular case of the static limit where the inflaton sits in the minimum, or equivalently the brane stands in the vacuum configuration. As a first, necessary, step in embedding the full dynamical process, we then looked for such a vacuum with the appropriate ingredients. As a warm-up, we first looked for a Minkowski supersymmetric vacuum in section \ref{sec:susy}, and could prove the absence of solutions on a particular internal compact manifold: the product of the nilmanifold $N_3$ required by the mechanism, times a three torus. For more general manifolds, we indicated generic difficulties to find a vacuum. We then turned in section \ref{sec:gensearch} to a more complete search for an appropriate vacuum using different tools than supersymmetry: essentially manipulations of supergravity equations of motion. This allowed us to prove a no-go theorem in section \ref{sec:nogo}, that concludes negatively on the existence of an appropriate vacuum. This means that, at least within the range of (reasonable) assumptions listed in section \ref{sec:nogo}, it is impossible to realise the monodromy inflation mechanism of \cite{Silverstein:2008sg} in a string compactification. We now make further comments.

\begin{itemize}
\item {\bf There exists no appropriate vacuum to embed the monodromy inflation mechanism of \cite{Silverstein:2008sg} in a concrete compactification setting.}
\end{itemize}

The no-go theorem obtained in section \ref{sec:nogo} excludes any anti-de Sitter or Minkowski vacuum (including supersymmetric ones). De Sitter vacua are not excluded, but have a lower bound on the cosmological constant: this bound, related to the curvature of the internal manifold, is too high for phenomenology. This excludes any realistic de Sitter vacuum. Note that de Sitter vacua are often obtained in practice by small deformations from anti-de Sitter or Minkowski ones, by considering non-perturbative corrections, or adding some flux, etc. Here, the absence of the last two vacua makes it even more unlikely to obtain an (unrealistic) de Sitter vacuum.

We believe that the absence of vacuum is essentially due to the direction wrapped by the brane together with the nature of the internal manifold: these two ingredients are constrained by the inflation mechanism to configurations that do not fit together to have a vacuum. This point is explained at the end of section \ref{sec:gensearchsusy} and was discussed in \cite{Andriot:2015sia}. In short, the brane wraps a internal subspace whose (smeared) world-volume form is closed: $\d \widetilde{{\rm vol}}_{||}=0$. A vacuum with such a brane is more likely to be found on a Ricci flat twisted torus (solvmanifold). However, the mechanism requires a three-dimensional part of the internal manifold to be given by the nilmanifold $N_3$, that is negatively curved. This contradiction can be traced in the final relation \eqref{rel2} that leads to the lower bound on the cosmological constant. The flux $F_4$ is the one sourced by the brane and encodes its properties. If the $\widetilde{{\rm vol}}_{||}$ was not closed, the flux would get an additional (constant) term that could compensate the internal curvature $\tilde{{\cal R}}_6$: this is precisely what happens for the vacuum given in section \ref{sec:N3T3}. Here instead, nothing compensates the internal curvature: this leads to a lower bound $-\tilde{{\cal R}}_{6_0}$ given by the curvature of $N_3$. Note that these two ingredients, preventing to obtain an appropriate vacuum, are topological requirements, that would not be altered by small deformations.

\begin{itemize}
\item {\bf Are there ways of circumventing the no-go theorem?}
\end{itemize}

By definition, circumventing the no-go theorem would require to violate one of the assumptions listed in section \ref{sec:nogo}. The simplest option would be to consider a $D_p$-brane and $O_p$-plane with $p\neq4$. An encouraging fact is that there are more supersymmetric Minkowski vacua known with $p=5$ or $p=6$. However, the mechanism has been designed precisely for $p=4$, so one should in the first place verify how to reproduce it for a different size of the sources, and rederive a suitable inflaton potential. In addition, the problem mentioned above on the directions wrapped may still be present: one may still require $\d \widetilde{{\rm vol}}_{||}=0$ for the mechanism to work. Vacua verifying this on solvmanifolds are rare and some were obtained in \cite{Andriot:2015sia} on Ricci flat solvmanifolds. Those (their fibration and metric) are however very different than the nilmanifold $N_3$ required by the mechanism. So in the end, changing the size of the sources may not ease the search for an appropriate vacuum to embed the inflation mechanism.

\begin{itemize}
\item {\bf What do we learn for cosmology and other proposed mechanisms?}
\end{itemize}

Many mechanisms, of inflation or even other scenarios, have been proposed to try to reproduce the recent cosmological observations. There are certainly motivations to have mechanisms in the context of string theory, but those often require a standard compactification setting in supergravity, to connect string theory to the four-dimensional physics. Obtaining this concrete compactification thus remains crucial. In that respect, the method followed in the present paper for the  monodromy inflation mechanism of \cite{Silverstein:2008sg} could be applied to other cases, namely the idea of determining the vacuum that includes the necessary ingredients for the mechanism to work. We have presented various tools and assumptions to find such an appropriate vacuum, and those could in principle be used for different mechanisms.

In the context of inflation mechanisms, the idea of using an axion for the inflaton and benefit this way from the shift symmetry has been very fruitful. An important contribution of the paper \cite{Silverstein:2008sg} has certainly been to emphasize the interest of axions having a monodromy; this lead to many ``axion monodromy'' mechanisms. For those, the difficulties stressed in the present paper to find an appropriate vacuum may also be useful, in particular the tension between having $\d \widetilde{{\rm vol}}_{||}=0$ and a curved underlying manifold. Let us give an example and consider the massive Wilson line axion monodromy mechanism of \cite{Marchesano:2014mla}. There, the inflaton is given by the v.e.v. $\varphi$ of a Wilson line on a brane in type II supergravity. In short, one has the flux ${\cal F}_2= \varphi\, \d \eta_1$ living by definition on the brane. The one-form $\eta_1$ should not be closed, and the context of twisted tori is then suited to obtain a concrete model, but $\d {\cal F}_2=0$. If we do not want to enter the difficulties related to codimension-two branes or less, this mechanism thus requires to focus on a $D_6$ or $D_5$ on twisted tori. Because ${\cal F}_2$ is along the brane, we are back for a $D_5$ to the condition $\d \widetilde{{\rm vol}}_{||}=0$. For a $D_6$, considering it to wrap e.g. a free circle times the two directions along ${\cal F}_2$ also gives $\d \widetilde{{\rm vol}}_{||}=0$, so it is not simple to avoid this condition.\footnote{In section 3.6.3 of \cite{Marchesano:2014mla}, the mechanism is exemplified with a probe $D_6$-brane wrapping a subspace with $\d \widetilde{{\rm vol}}_{||}=0$. Although no parallel orientifold is proposed to cancel the resulting tadpole, considering one there would be allowed. The presence of other intersecting branes and planes in the background makes it however very difficult to localize, i.e. backreact, the probe $D_6$-brane and $O_6$-plane, making them effectively absent of the vacuum.} As pointed-out, it is however difficult to get vacua on twisted tori with branes verifying $\d \widetilde{{\rm vol}}_{||}=0$. So the difficulties stressed in the present paper may also apply and constrain generally other axion monodromy inflation mechanisms. It would still be interesting to use the vacua of \cite{Andriot:2015sia} to try to embed this massive Wilson line axion monodromy mechanism.

\vspace{0.4in}

\subsection*{Acknowledgements}

I would like to thank D. Junghans and G. Shiu for very helpful discussions, in particular at an early stage of this project when they were involved. I also thank Y. Korovin and A. Tomasiello for useful related discussions. This work is part of the Einstein Research Project ``Gravitation and High Energy Physics'', which is funded by the Einstein Foundation Berlin.

\newpage

\providecommand{\href}[2]{#2}\begingroup\raggedright

\endgroup

\end{document}